\documentstyle[12pt]{article}

\catcode`\@=11
\long\def\@makefntext#1{
\protect\noindent \hbox to 3.2pt {\hskip-.9pt
$^{{\ninerm\@thefnmark}}$\hfil}#1\hfill}		

\def\@makefnmark{\hbox to 0pt{$^{\@thefnmark}$\hss}}  

\def\ps@myheadings{\let\@mkboth\@gobbletwo
\def\@oddhead{\hbox{}
\rightmark\hfil\ninerm\thepage}
\def\@oddfoot{}\def\@evenhead{\ninerm\thepage\hfil
\leftmark\hbox{}}\def\@evenfoot{}
\def\sectionmark##1{}\def\subsectionmark##1{}}

\renewcommand{\thefootnote}{\fnsymbol{footnote}}

\newcounter{sectionc}\newcounter{subsectionc}\newcounter{subsubsectionc}
\renewcommand{\section}[1] {\vspace*{0.6cm}\addtocounter{sectionc}{1}
\setcounter{subsectionc}{0}\setcounter{subsubsectionc}{0}\noindent
	{\normalsize\bf\thesectionc. #1}\par\vspace*{0.4cm}}
\renewcommand{\subsection}[1] {\vspace*{0.6cm}\addtocounter{subsectionc}{1}
	\setcounter{subsubsectionc}{0}\noindent
	{\normalsize\it\thesectionc.\thesubsectionc. #1}\par\vspace*{0.4cm}}
\renewcommand{\subsubsection}[1]
{\vspace*{0.6cm}\addtocounter{subsubsectionc}{1}
	\noindent {\normalsize\rm\thesectionc.\thesubsectionc.\thesubsubsectionc.
	#1}\par\vspace*{0.4cm}}

\newcounter{appendixc}
\newcounter{subappendixc}[appendixc]
\newcounter{subsubappendixc}[subappendixc]

\renewcommand{\appendix}[1] {\vspace*{0.6cm}
        \refstepcounter{appendixc}
        \setcounter{figure}{0}
        \setcounter{table}{0}
        \setcounter{equation}{0}
        \renewcommand{\thefigure}{\Alph{appendixc}.\arabic{figure}}
        \renewcommand{\thetable}{\Alph{appendixc}.\arabic{table}}
        \renewcommand{\theappendixc}{\Alph{appendixc}}
        \renewcommand{\theequation}{\Alph{appendixc}.\arabic{equation}}
        \noindent{\bf Appendix \theappendixc #1}\par\vspace*{0.4cm}}

\def\abstracts#1{{
	\centering{\begin{minipage}{12.2truecm}\footnotesize\baselineskip=12pt\noindent
	\centerline{\footnotesize ABSTRACT}\vspace*{0.3cm}
	\parindent=0pt #1
	\end{minipage}}\par}}

\renewenvironment{thebibliography}[1]
	{\begin{list}{\arabic{enumi}.}
	{\usecounter{enumi}\setlength{\parsep}{0pt}
\setlength{\leftmargin 1.25cm}{\rightmargin 0pt}
	 \setlength{\itemsep}{0pt} \settowidth
	{\labelwidth}{#1.}\sloppy}}{\end{list}}

\newcounter{itemlistc}
\newcounter{romanlistc}
\newcounter{alphlistc}
\newcounter{arabiclistc}

\newcommand{\fcaption}[1]{
        \refstepcounter{figure}
        \setbox\@tempboxa = \hbox{\footnotesize Fig.~\thefigure. #1}
        \ifdim \wd\@tempboxa > 6in
           {\begin{center}
        \parbox{6in}{\footnotesize\baselineskip=12pt Fig.~\thefigure. #1}
            \end{center}}
        \else
             {\begin{center}
             {\footnotesize Fig.~\thefigure. #1}
              \end{center}}
        \fi}

\newcommand{\tcaption}[1]{
        \refstepcounter{table}
        \setbox\@tempboxa = \hbox{\footnotesize Table~\thetable. #1}
        \ifdim \wd\@tempboxa > 6in
           {\begin{center}
        \parbox{6in}{\footnotesize\baselineskip=12pt Table~\thetable. #1}
            \end{center}}
        \else
             {\begin{center}
             {\footnotesize Table~\thetable. #1}
              \end{center}}
        \fi}

\def\@citex[#1]#2{\if@filesw\immediate\write\@auxout
	{\string\citation{#2}}\fi
\def\@citea{}\@cite{\@for\@citeb:=#2\do
	{\@citea\def\@citea{,}\@ifundefined
	{b@\@citeb}{{\bf ?}\@warning
	{Citation `\@citeb' on page \thepage \space undefined}}
	{\csname b@\@citeb\endcsname}}}{#1}}

\newif\if@cghi
\def\cite{\@cghitrue\@ifnextchar [{\@tempswatrue
	\@citex}{\@tempswafalse\@citex[]}}
\def\citelow{\@cghifalse\@ifnextchar [{\@tempswatrue
	\@citex}{\@tempswafalse\@citex[]}}
\def\@cite#1#2{{$\null^{#1}$\if@tempswa\typeout
	{IJCGA warning: optional citation argument
	ignored: `#2'} \fi}}

 1
 1
 1

\font\ninerm=cmr9

\begin{document}

\centerline{\normalsize\bf Theoretical Response to the Discovery of}
\baselineskip=16pt
\centerline{\normalsize\bf Deeply Bound Pionic States 
in $^{208}$Pb(d, $^3$He) Reactions}
\baselineskip=22pt

\vspace*{0.6cm}
\centerline{\footnotesize H. Toki$^{a)}$, S. Hirenzaki$^{b)}$ and
K. Takahashi$^{b)}$}
\baselineskip=16pt
\centerline{\footnotesize\it a) RCNP, Osaka University, Ibaraki, Osaka 567,
Japan}
\baselineskip=12pt
\centerline{\footnotesize\it b) Nara Women's University, Nara 630, Japan}

\vspace*{0.9cm}
\abstracts{Recently, deeply bound pionic states were found experimentally 
in (d, $^3$He) reactions on $^{208}$Pb. They found an isolated peak structure in
the bound region below the pion production threshold. We study theoretically
these excitation functions in (d, $^3$He) reactions on $^{208}$Pb at
T$_d$=600 MeV. We found very good agreement with the (d, $^3$He) excitation
functions and could identify the underlying structures of the pionic states.
We study the energy dependence of the (d, $^3$He) reactions and the change of
the excitation functions with the incident energy.}

\vspace*{0.8cm}
\normalsize\baselineskip=15pt
\setcounter{footnote}{0}
\renewcommand{\thefootnote}{\alph{footnote}}

The existence of deeply bound pionic atoms in heavy nuclei was predicted 
by Toki and Yamazaki as
quasi-static states$^{1),2)}$. Since then many efforts were made for detection
of these states both experimentally and theoretically$^{3)-10)}$. In particular,
it is worth mentioning that (n,d) reactions$^{9)}$ and (p,pp) reactions$^{10)}$ 
on the
$^{208}$Pb target were able to identify some strength in the excitation
function below the pion production threshold. It is, however, not yet
convincing due to the lack of good resolution and good statistical accuracy. 
It was, hence, badly needed to
perform the similar experiments with better resolution and statistical
accuracy for identification of
isolated peak structure in the bound region below the pion production
threshold.

Very recently, Yamazaki et.al. performed such experiments with the use of (d,
$^3$He) reactions on $^{208}$Pb with better resolution and statistical 
accuracy$^{11)}$. They, at last, succeeded
to identify clearly a peak structure in the bound region. The results were
presented in the recent publication with the theoretical predictions, which
were made before the experiments were performed$^{6)}$. The agreement of the
theoretical predictions to the newly obtained experimental data is 
almost perfect.

In this paper, we would like to describe in detail the theoretical part of the
(d, $^3$He) reaction on $^{208}$Pb. We should describe the underlying structures
of the (d, $^3$He) excitation functions. At the same time, we would like to
show the spectrum with better resolution to predict what should be expected 
when
experiments were performed with better (feasible) resolution. In addition, we
would like to show the energy dependence of the excitation function of the (d,
$^3$He) reaction with the $^{208}$Pb target.

We describe the (d, $^3$He) reaction within the effective number approach. The
(d, $^3$He) cross sections use the experimental cross section and the 
effective number of nucleons exciting a pion in a definite orbit with 
inclusion of the distortion effect. 
The (d, $^3$He) cross section in the laboratory frame is expressed as$^{4)}$


$${\left({{d\,\rm \sigma  \over \rm \mit d\,\rm \Omega }}\right)}_{d\ +\ A\ \rm 
\rightarrow \rm \ ^{3} \it He\rm \pi }\ =\ {\left({{d\,\rm \sigma  \over \rm
\mit d\,\rm \Omega }}\right)}_{dn\ \rm  \rightarrow \rm \ ^{3} \it He\rm \pi
}^{lab}{\ N}_{eff}$$
Here, the effective number takes care of the initial neutron in a shell model
orbit specified by $j_n$ and the final pion in a pionic atom orbit specified by
$l_\pi$ with the initial deuteron and final $^3$He distortions. After
manipulations, we find $N_{eff}$ for the configuration $[l_{\pi} \otimes 
j_n^{-1}]J$ be written in the following form

$${N}_{eff}\ ={1 \over 2}\rm \sum\limits_{\rm \mit Lm}^{} {\left|{{\rm \mit
S}_{\rm \mit JL}\int_{}^{}\rm \mit \exp\ (i\vec{q}\vec{r})\ D(b)\ 
{R}_{\rm \pi}(r)\ {R}_{n}(r){Y}_{Lm}(\widehat{r}){d}^{3}r}\right|}
^{\rm \mit 2}$$
with the statistical factor

$$\matrix{{S}_{JL}\ =\ \left\langle{{l}_{\rm \pi }({l}_{\rm \pi }\ {1 \over
2})jn;\ J\ |\ ({l}_{\rm \pi }\ {l}_{n})L{1 \over 2};\ J}\right\rangle\sqrt
{{2J\ \ +\ 1 \over 2L\ +\ 1}}{(-)}^{{l}_{n}}\cr
\rm \times \rm \mit \ \sqrt {{(2l\rm \pi \rm \mit \ +1)(2\ln\ +1) \over 4\rm
\pi \rm \mit (2L\ +1)}}\ ({l}_{\rm \pi }o{\ l}_{n}o\ |\ o)\cr}$$
The distortion factor D(b) is estimated by the eikonal approximation,

$$D(b)\ =\ \exp\ (-\ {1 \over 2}\int_{-\rm  \infty }^{\rm \infty }
\overline{\rm \sigma }\rho(z',\ b)dz')$$
As for the average cross section $\bar{\sigma}$, we take $\bar{\sigma}=
\frac{1}{2}(\sigma_{dN}+\sigma_{^3HeN})$, where N denotes nucleon. 
We calculate the cross section leading to a pion in the continuum 
in the similar way 
as for the bound states. Essentially, we replace the bound pion wave 
function by a scattering pion wave function, obtained by solving the 
distorted wave optical model. All the details will be published in the near
future$^{12)}$.

The important information for the pionic atom formation is the elementary
differential cross sections at the forward angle for $d+n \rightarrow \
^3He+\pi^-$ in the laboratory frame. We first discuss the incident energy
dependence of the elementary differential cross sections, which are shown 
in Fig.3 of Ref. 4. The cross section peaks around
T$_d$ = 600 MeV. The momentum transfers for (d, $^3He$) reactions are provided
in Fig.2 of Ref.4. At T$_d$ = 600 MeV, the momentum transfer is small. This
small momentum transfer and the large cross section at T$_d$ = 600 MeV seem
ideal for the study of deeply bound pionic states in heavy nuclei.

We present here the numerical results for $^{208}$Pb(d, $^3$He) reactions for
pionic atom formation. As for the bound neutron, we take two major shells below
the Fermi surface; $50<N<126$. The separation energies of the neutron orbitals
are $s_n$ = 7.367 MeV for $p_{1/2}$, $s_n$ = 7.937 MeV for $f_{5/2}$, $S_n$ =
8.264 MeV for $p_{3/2}$ and $S_n$ = 9.000 MeV for $i_{13/2}$. The pionic states
are calculated by using the optical potential of Seki and Masutani$^{13)}$ and
the energies and the widths are given in Ref.2.

We show the calculated results on $^{208}$Pb(d, $^3$He) reactions at
T$_d$ = 600 MeV in Fig.1. In the upper part of Fig.1, we show the forward cross
sections with the resolution of 500 keV in order to compare with the
experimental spectra shown in the middle part of Fig.1. The agreement is almost
perfect. We note that we have added the contributions from deeper neutron
single particle orbits $(50<N<82)$ in the calculation of the continuum cross
sections as compared to the one in the discovery paper$^{11)}$. We assume
20$\mu$b/sr/MeV constant background in the theoretical spectra.

In order to see the underling structures and also the spectra with a better
experimental resolution, we show in the lower part of Fig.1, the cross sections
with 200 keV resolution. We can see now many peaks below the pion production
threshold, which is denoted by the vertical dashed line.
The peak seen experimentally consists of two states; $[l_\pi \otimes
j_n^{-1}]=[1 \otimes p_{1/2}^{-1}]$ and $[1 \otimes p_{3/2}^{-1}]$. Since 4
neutrons are allowed in the $p_{3/2}$ state and 2 neutrons in the $p_{1/2}$
state, the ratio of the cross sections leading to $p_{3/2}^{-1}$ and 
$p_{1/2}^{-1}$ is 2. 
Those two states provide the peak and the shoulder structure at 
Q = -136 MeV (around 5 MeV binding energy).
The pionic $s$ state seems to be only weakly excited due to the necessity
of some momentum transfer. 
In addition, there is an appreciable amount of contribution from the
neutron $f_{5/2}$ hole coupled with the pionic $s$ state to this small peak.
Many more peaks are seen closer to the pion
production threshold. The dominant contributions are again coming from the
pionic atom states coupled with the $p_{3/2}$ and $p_{1/2}$ neutron holes.
It is interesting to comment on the bump structure seen in the calculated
spectra around Q = -143 MeV in the continuum. This bump structure is caused by
exciting $[1s_\pi \otimes s_{1/2}^{-1}]$ state, where a $s_{1/2}$ neutron in
the $50<N<82$ major shell is picked up and a pion is placed in the $1s$ orbit.
In the calculation, the width of the $s_{1/2}$ neutron hole state is assumed
zero.

We study now the energy dependence of the (d, $^3$He) cross sections. As could
be guessed from the energy dependence of the elementary differential cross
sections and the larger momentum transfers, the (d, $^3$He) cross
sections at different energies should decrease. It is, however, interesting to
see how the cross sections will change with the incident energy. For this
purpose, we show the calculated results at T$_d$ = 500, 800 and 1000 MeV with
the resolution of $\Delta$E = 200 keV in Fig.2.

When we lower incident energy to T$_d$ = 500 MeV, the peak structure at E
$\sim$
5 MeV is enhanced relative to other structure due to smaller momentum transfer.
The peak structures are almost exclusively produced by the $p$ neutron hole
contribution. Note that the 1$s$ pionic state is not appreciably excited at
this energy. Those energy dependences are clearly explained by using the energy
dependence figures; Fig.5-8 in Ref.4.
As the energy is increased to T$_d$ = 800 MeV, the cross sections
are lowered. As indicated by the three kinds of curves, now the contributions
from the $p$ neutron hole state become small, while the contributions from the
$i_{13/2}$ and $f_{5/2}$ neutron hole states increase. 
Pionic states coupled with $f_{5/2}$ neutron hole state are excited
appreciably.
Further increased to T$_d$ = 1000 MeV,
the peak structures are completely dominated by the $i_{13/2}$
neutron hole state coupled with various pionic atom states.

It is amazing to observe that the calculated peak structures agree extremely
well with the experimental results. Since various pionic atom - neutron hole
states are excited by the (d, $^3$He) reactions, we expect some pion-hole
residual interactions to mix these states. We ought to estimate this mechanism
theoretically. We believe, however, that the residual interactions are much
smaller than the case of nuclear particle hole states, since the overlaps
between pionic states and the nuclear states are quite small.
 
We have studied theoretically the (d, $^3$He) reactions leading to pionic atom
states in $^{208}$Pb at T$_d$ = 600 MeV as the theoretical response to the
recent discovery of deeply bound pionic atoms $^{11)}$. 
The peak structure seen around
E $\sim$ 5 MeV is due to 2$p$ pionic state coupled with p$_{3/2}$ and p$_{1/2}$
neutron hole states. We have shown the excitation function of (d, $^3$He)
reactions at the same incident energy with a better resolution, 
$\Delta$E = 200 keV, to see the peak structures clearly. In addition, 
we have presented the
energy dependence of the excitation functions at various incident energies. At
higher energy as T$_d$ = 1000 MeV, the $i_{13/2}$ neutron hole state coupled
with various pionic atom states dominate the excitation function.

We are greatful to Prof. T. Yamazaki for communicating the experimental
results much before its publication and also for various enlightening
discussions. We thank also R. Hayano, K. Itahashi and A. Gillitzer for 
discussions and sending us the detailed numbers.

\pagebreak

Fig.1
Forward cross sections of the $^{208}$Pb(d, $^3$He) reaction at T$_d$ = 600
MeV; (a) Theoretical results with the experimental resolution of 500 keV, (b)
Observed spectrum by Yamazaki et.al.$^{11)}$, and (c) Theoretical results with
200 keV experimental resolution. In (a) and (c) the flat background is assumed
to be 20 $\mu$b/sr/MeV. The vertical dashed line denotes the $\pi^-$ emission
threshold energy.

\vspace*{0.3cm}
Fig.2
Calculated cross sections of the $^{208}$Pb(d, $^3$He) reaction at (a) T$_d$ =
500 MeV, (b) T$_d$ = 800 MeV, and (c) T$_d$ = 1000 MeV with 200 keV
experimental energy resolution. Solid curves show the full spectra including
all contributions. Dotted curves show the contributions from the neutron 2$p$
states ($p_{3/2}$ and $p_{1/2}$) and dashed curve show those from the neutron
$i_{13/2}$ state. Added is also the contribution from the neutron $1f$ states
 ($f_{5/2}$ and $f_{7/2}$) by dash-dotted curve in the middle figure.


\end{document}